\documentclass[11pt]{article}

\usepackage{amsmath}
\usepackage{amssymb}
\usepackage{mathptmx}
\usepackage{amsfonts}
\usepackage[mathscr]{eucal}
\usepackage[dvips]{graphicx}
\usepackage{color}

\newcommand{\mbf}[1]{\ensuremath{\mathbf{#1}}}
\newcommand{\ms}[1]{\ensuremath{\mathscr{#1}}}

\newcommand{\sqzb}[1]{\ensuremath{d\Gamma_{\textrm{b}}({#1}) }}
\newcommand{\sqzf}[1]{\ensuremath{d\Gamma_{\textrm{f}}({#1}) }}
\newcommand{\domain}[1]{\ensuremath{\mathscr{D} ({#1})}}

\newcommand{\tens}{\otimes}

\newcommand{\nstens}{\otimes^{n}_{s}}
\newcommand{\natens}{\otimes^{n}_{a}}

\newcommand{\piE}{( 2 \pi )^{3}E_{M} }
\newcommand{\eltwo}{L^{2}(\mathbf{R}^{3})}
\newcommand{\Rthree}{\mathbf{R}^{3} }
\newcommand{\dx}{d \mathbf{x} }
\newcommand{\dy}{d \mathbf{y} }
\newcommand{\dz}{d \mathbf{z} }
\newcommand{\dk}{d \mathbf{k} }

\newcommand{\core}{\ms{F}^{\textrm{fin}}_{\textrm{Dirac}} (\ms{D}(E_{M}))
 \hat{\otimes} \ms{F}^{\textrm{fin}}_{\textrm{rad}}  (\ms{D}(\omega))}

\newcommand{\I}{\textrm{I}}
\newcommand{\II}{\textrm{II}}

\newcommand{\fin}{\textrm{fin}}
\newcommand{\Dzero}{ \mathscr{D}_{0} }
\newcommand{\QED}{\textrm{QED}}
\newcommand{\Dirac}{ \textrm{Dirac} }
\newcommand{\rad}{ \textrm{rad} }
\newcommand{\R}{ \textrm{R} }
\newcommand{\Dir}{ \textrm{Dir} }
\newcommand{\D}{ \textrm{D} }
\newcommand{\spa}{ {} }
\newcommand{\longti}{ \textrm{long} }
\newcommand{\HDirac}{H_{\textrm{Dirac}}}
\newcommand{\Hrad}{H_{\textrm{rad}}}
\newcommand{\Hlongti}{H_{\textrm{long}}}
\newcommand{\Hzero}{H_{0}}
\newcommand{\Veff}{V_{\textrm{eff}}}

\newcommand{\POmegarad}{P_{\Omega_{\textrm{rad}}}}
\newcommand{\Omegarad}{\Omega_{\textrm{rad}}}
\newcommand{\btriangle}{\bigtriangleup}

\newcommand{\density}[1]{\ensuremath{\rho{(\textbf{#1})}}}
\newcommand{\current}[2]{\ensuremath{
 J^{#2} (\mathbf{#1}) }}
\newcommand{\cutoff}[2]{\ensuremath{
\chi_{\textrm{#2}} (\mathbf{#1} )}}
\newcommand{\opeA}[2]{\ensuremath{
 A^{#2} (\mathbf{#1} ) }}
\newcommand{\opePi}[2]{\ensuremath{
 \Pi^{#2} (\mathbf{#1} ) }}

\newcommand{\stateRad}{L^{2}( \mathbf{R}^{3} ; \mathbf{C}^{2} )}
\newcommand{\stateDirac}{L^{2}( \mathbf{R}^{3} ; \mathbf{C}^{4} )}
\newcommand{\intRthree}{\int_{\mathbf{R}^{3}}}
\newcommand{\FQED}{\mathscr{F}_{\textrm{QED}}}

\newtheorem{theorem}{Theorem}[section]
\newtheorem{proposition}[theorem]{Proposition}
\newtheorem{lemma}[theorem]{Lemma}
\newtheorem{corollary}[theorem]{Corollary}

\begin{document}

\title{Scaling Limit of   Quantum  
 Electrodynamics  \\
 with  Spatial Cutoffs   
}
\author{Toshimitsu TAKAESU}
\date{ }
\maketitle
\begin{center}
\textit{Faculty of Mathematics, Kyushu University, \\
 Fukuoka, 812-8581, Japan }
\end{center}

\begin{quote}
\textbf{Abstract.}
 In this paper  the Hamiltonian of  quantum electrodynamics with spatial cutoffs is investigated.
 We define a scaled total Hamiltonian and consider its asymptotic behavior.  
 In the main theorem,  it is shown that the scaled total Hamiltonian converges to a self-adjoint operator in the strong resolvent sense, and
   effective potentials are derived.         
\end{quote}

\section{Introduction}
$\; $ 
Quantum electrodynamics (QED)  describes the system 
of  Dirac fields coupled to quantized 
radiation fields.  In this paper a scaled QED Hamiltonian is considered. In the main theorem,  the effective potentials  are obtained by taking a  scaling limit of the scaled QED Hamiltonian. 
Let us define a QED Hamiltonian as an operator on a boson-fermion Fock space. 
The state space of QED is defined by the boson-fermion Fock space   $\ms{F}_{\QED}  = \ms{F}_{\Dirac} \tens \ms{F}_{\rad} \; $, where $\ms{F}_{\Dirac}$ is the fermion Fock space on $\stateDirac$ and $\ms{F}_{\rad}$ is the boson Fock space
 on $\stateRad$. The field operators of the Dirac field and the radiation field are denoted by
 $\psi (\mbf{x}) $ and $\mbf{A} (\mbf{x})$, respectively. Here  we impose
  ultraviolet cutoffs on both $\psi (\mbf{x}) $ and $\mbf{A} (\mbf{x})$.
 To define a interaction between the Dirac field and the radiation field,  we introduce the electromagnetic current : 
 \[
   J (\mbf{x}) \; \; = \; \; 
\left[
\begin{array}{c}
\rho (\mbf{x}) \\ \mbf{J} (\mbf{x})
\end{array}
\right] , 
\]
where $\; \rho(\mbf{x}) =\psi^{\ast} (\mbf{x}) \psi(\mbf{x})  $ and  $\; J^j (\mbf{x}) = \psi^{\ast} (\mbf{x})
 \alpha^j \psi (\mbf{x})  $,  $\; j=1, 2, 3$, with $ \alpha^j \in M_{4}(\mbf{C})$ satisfying the canonical anti-commutation relation $\{ \alpha^j  , \alpha^l \} = 2 \delta_{j,l}$. In this paper, instead of $J (\mbf{x})$,  we consider the spatially localized electromagnetic current :
  \[
   J_{\chi} (\mbf{x}) \; \; = \; \; 
\left[
\begin{array}{c}
 \rho_{\chi} (\mbf{x}) \\  \mbf{J}_{\chi} (\mbf{x})
\end{array}
\right] , 
\]
 where $ \rho_{\chi} (\mbf{x}) = \chi (\mbf{x})  \rho (\mbf{x}) $ and  $\; \mbf{J}_{\chi} (\mbf{x})= \chi (\mbf{x}) \mbf{J} (\mbf{x})$  with a spatial cutoff $\chi (\mbf{x})$.
 Then the QED Hamiltonian  with the spatial cutoff  is 
defined by 
 \begin{equation}
 H \; \; = \; \;  \HDirac  + \Hrad  + \; 
 e \intRthree  
  \mbf{J}_{\chi} (\mbf{x}) \mbf{\cdot} \mbf{A} (\mbf{x} ) d\mbf{x} \; 
+ \; 
 \frac{e^2}{8 \pi} \int_{\Rthree \times \Rthree}
\frac{ \rho_{\chi} (\mbf{x}) \rho_{\chi} (\mbf{y}) }{|\mbf{x} -\mathbf{y}|}
      d \mbf{x} \, d\mathbf{y} ,
\end{equation} 
where $\HDirac $ and $\Hrad$ are the free Hamiltonians of the Dirac field and the radiation field, respectively,
 $\; e \in \mbf{R}\; $ denotes the coupling constant, and  $\mbf{J}_{\chi} (\mbf{x}) \mbf{\cdot} \mbf{A} (\mbf{x} )= \sum\limits_{j=1}^{3}  J^{j}_{\chi} (\mbf{x}) A^{j}(\mbf{x} )$. $\HDirac $ and $\Hrad$ are denoted  by formally 
\begin{align*}
&\HDirac \; = \sum_{s=\pm 1/2}\int_{\Rthree}
 \sqrt{\mbf{p}^2 + M^2}  \left( \frac{}{}
  b_{s}^{\ast} (\mbf{p}) b_{s} (\mbf{p}) \; + \;  d_{s}^{\ast} (\mbf{p}  ) d_{s} (\mbf{p}) \right) d \mbf{p} ,  \qquad \qquad M > 0 , \\
  & \Hrad \; = \sum_{r=1,2}\int_{\Rthree}
   | \mbf{k}| a_{r}^{\ast}(\mbf{k})  a_{r}(\mbf{k}) \dk  .
\end{align*}
It is seen that 
 under some conditions on ultraviolet cutoffs and spatial cutoffs, 
  $H$ is a self-adjoint operator on  
 $\ms{F}_{\QED} $ in  \cite{Ta09}, and the spectral properties of $H$ also has been investigated in 
\cite{BDG04, DiGu03, Ta09}.

$\quad  $ \\ 
Now we consider the scaled QED Hamiltonian defined by
 \begin{equation}
 H (\Lambda )\; \; = \; \;  \HDirac  + \Lambda^2 \Hrad  + \; 
 e \Lambda \intRthree  
  \mbf{J}_{\chi} (\mbf{x}) \mbf{\cdot} \mbf{A} (\mbf{x} ) d\mbf{x} \; 
+ \; 
 \frac{e^2}{8 \pi} \int_{\Rthree \times \Rthree}
\frac{ \rho_{\chi} (\mbf{x}) \rho_{\chi} (\mbf{y}) }{|\mbf{x} -\mathbf{y}|}
      d \mbf{x} \, d\mathbf{y} ,  \; \; \Lambda > 0 ,\label{scaledH}
\end{equation} 
 and  this  is the main object in this paper. 
 Historically Davies investigates a scaled Hamiltonian of the form  $H_{\textrm{p}}+\Lambda \kappa \phi (\mbf{x})+ \Lambda ^2  H_{\textrm{b}}$ in \cite{Da79} where $H_{\textrm{p}}=\frac{\mbf{p}^2}{2M}$ is a shcr\"{o}dinger operator, $\phi (\mbf{x}) $ is the field operator of the scalar bose field, and  $H_{\textrm{b}} $ is the free Hamiltonian. Then  an effective Hamiltonian $H_{\textrm{p}}+\kappa^2 
V_{\textrm{eff} }(\mbf{x})$ is obtained by the scaling limit of the scaled Hamiltonian. This is the so called weak coupling limit. Regarding this scaling limit as  $
\exp (-it\Lambda^2 (\Lambda^{-2} H_{\textrm{p}}+\Lambda^{-1} \kappa \phi  (\mbf{x})+ H_{\textrm{b}})  $,
we may say that the weak coupling limit is to take $t\to \infty$, $M \to \infty$ and $\kappa \to0$ simultaneously. Roughly speaking it is a long time behavior of the time evolution of the Hamiltonian but with a simultaneous weak coupling limit between a particle and a scalar field. The scaled QED Hamiltonian $H (\Lambda) $ in  (\ref{scaledH}) is an extended  model  consided in Davies \cite{Da79}, and the unitary evolution of 
$H (\Lambda ) $ is given by 
  {\large
\begin{equation}
 e^{-it H (\Lambda ) } \; \; = \; \; e^{-it\Lambda^2  \left( 
\frac{1}{\Lambda^2}\HDirac  + \Hrad  + \; 
 \left( \frac{e}{\Lambda} \right)  \intRthree  
  \mbf{J}_{\chi} (\mbf{x}) \mbf{\cdot} \mbf{A} (\mbf{x} ) d\mbf{x} \; 
+ \;  \frac{1}{8 \pi}
 \left( \frac{e}{\Lambda} \right)^2 \int_{\Rthree \times \Rthree}
\frac{ \rho_{\chi} (\mbf{x}) \rho_{\chi} (\mbf{y}) }{|\mbf{x} -\mathbf{y}|}
      d \mbf{x} \, d\mathbf{y}  \right)} , 
\end{equation} 
}
where  $t \Lambda^2 $ is the scaled time and $\frac{e}{\Lambda}$ is the scaled coupling constant.
As a remark, $H(\Lambda )$ is  also derived from the transformation 
$a_{r} (\mbf{k}) \; \mapsto  \; 
\Lambda  \,  a_{r}(\mbf{k})$, $\; a_{r}^{\ast} (\mbf{k}) \; \mapsto \;  
\Lambda \,  a_{r}^{\ast} (\mbf{k}) $. In this case, however,  the ultraviolet cutoffs are independent of the scaling parameter $\Lambda$.

$\quad$ \\
In the  main theorem,  the asymptotic behavior of $H (\Lambda )$ as $\Lambda \to \infty$ is considered.
 To investigate it,  we consider a 
 dressing  transformation, which is a unitary transformation,   defined in (\ref{HzeroK}).   Then  
by taking the   scaling limit of $H(\Lambda )  $ as $\Lambda \to \infty$, we have
\begin{equation}
s-\lim_{\Lambda \to \infty }
\left( H (\Lambda ) -z \frac{}{} \right)^{-1} \; \\  = 
\left( \HDirac + \frac{e^2}{8 \pi } \int_{\Rthree \times \Rthree}
\frac{\rho_{\chi} (\mbf{x}) \rho_{\chi} (\mbf{y}) }{|\mbf{x} -\mathbf{y}|}
      d \mbf{x} \, d\mathbf{y}   \;  -  \frac{e^2}{4} \Veff \;   -z \right)^{-1}
 \POmegarad ,  \label{6/23.1}
\end{equation}
where $V_{\textrm{eff}}$ is a effective potential of the Dirac field given by
\begin{equation}
\Veff \; = \; \int_{\Rthree \times \Rthree}
 \; \;  \mbf{J}_{\chi} (\mbf{x}) \cdot \triangle (\mbf{x}-\mbf{y})
  \mbf{J}_{\chi} (\mbf{y} )  \; \; \dx \dy ,
\end{equation}
 $ \triangle (\mbf{z} ) = ( \lambda^{j,l} (\mbf{z} ) + 
 \lambda^{j,l} (\mbf{-z} )  )_{j,l} $ is  $3 \times 3$ matrix with
 $ \lambda^{j,l} (\mbf{z} )$ defined in  
 (\ref{lambdajl}), and   
 $\POmegarad $ denotes the projection onto 
 the linear subspace spanned  by the Fock vacuum $\Omega_{\rad} \; \in \; \ms{F}_{\rad}$.
It is noted that  $\Veff $ is an operator on $\ms{F}_{\Dirac}$.  
Thus, by the scaling limit, we see formally that  the density of the charge is changed 
as follows : 
\[
\density{x} \density{y} \quad 
\longmapsto   \quad  \density{x} \density{y} \; \; + \; \; \textrm{const.}
 \;  |\mbf{x} -\mbf{y}|  \; \mbf{J} (\mbf{x}) \cdot \triangle (\mbf{x}-\mbf{y})
  \mbf{J} (\mbf{y} ) .
\]

$\quad $ \\
There are a lot of results  on 
 scaling limits  of  quantum field Hamiltonians, so far. 
As is mentioned above, the first rigorous result  is obtained  by
 Davies \cite{Da79}, and  he derives   $N$-body
 Sch\"{o}dinger Hamiltonians with  effective potentials 
 from the Hamiltonians of the system of particles interacting with   bose fields.   
Arai \cite{Ar90} considers an abstract scaling limit, and then
apply it to a spin-boson model and the non-relativistic QED models  in the dipole approximation.  For further  results on  the non-relativistic QED models,  refer to 
 \cite{Hi93, Hi97, Hi02, HiSp01}.
Hiroshima considers the Hamiltonian of a system of 
 particles coupled to Klein-Gordon fields \cite{Hi98, Hi99}. In  these papers,  Hiroshima  takes the scaling limit of the Hamiltonian, and  removes   ultraviolet cutoffs \textit{simultaneously}. Then he derives the Yukawa potential as an effective potential.  
 On the recent  research,  
Suzuki considers  generalized spin-boson model \cite{Su07-1} and generalized Nelson model \cite{Su07-2}, 
and Ohkubo investigates the so called Derezi\'{n}ski-G\'{e}rard
 model  \cite{Oh09}. 
 
$\quad$ \\
This paper is organized as follows. In Section 2, we introduce the  Dirac field and the quantized radiation field with ultraviolet cutoffs, and by introducing  spatial cutoffs, we  
 define the QED Hamiltonian  on the boson-fermion Fock space $\, \ms{F}_{\QED}  = \ms{F}_{\Dirac} \tens \ms{F}_{\rad} $, and state the main   theorem. In Section 3, we give the proof of the main theorem.

\section{Definitions and Main Results}
\subsection{Dirac Fields}
Let us first define the Dirac field \cite{Tha}.
The state space of the Dirac field is defined by
\begin{equation}
\ms{F}_{\Dirac} = \oplus_{n=0}^{\infty}
 (  \natens \stateDirac ) ,  \notag 
\end{equation}
where $\natens $ denotes the $n$-fold anti-symmetric tensor product with 
$ \tens_{a}^{0} L^{2}(\Rthree ; \mbf{C}^{4} ) := \mbf{C}$.
For $\xi ={}^{t} (\xi_{1}, \cdots , \xi_{4} )  \in L^{2}(\Rthree ; \mbf{C}^{4} )\; $,  we denote the annihilation operator by $B (\xi )$,  and for $\eta ={}^{t} (\eta_{1}, \cdots , \eta_{4} )  \in L^{2}(\Rthree ; \mbf{C}^{4} )\; $ the creation operator $B^{\ast} (
\eta )$.
The creation operators and annihilation operators  
 satisfy the canonical anti-commutation relations  : 
 \[
 \{ B (\xi ) , B^{\ast} (\eta ) \} \; = \; 
 ( \xi , \eta )_{L^{2}(\Rthree ; \mbf{C}^{4} )} ,
  \qquad  \{ B (\xi ) , B (\eta ) \} \; = \; 0 ,
 \]
  where $\{X , Y \} = XY +  YX $.
  In this paper the inner product $(y,x )_{\ms{H}}$ on a Hilbert space $\ms{H}$ is linear in  $x$ and antilinear in $y$. 
  Let $\Omega_{\Dir} = \{ 1, 0, 0, \cdots \} \; \in 
 \ms{F}_{\Dirac}  $ be the Fock vacuum.  The finite particle subspace over $\ms{N} 
 \subset \stateDirac $  is defined by
 \begin{equation}
\ms{F}_{\Dir}^{\fin}(\ms{N} )= 
\textrm{L.h} \{  B^{\ast} (\xi_{1}) \cdots  
B^{\ast} (\xi_{n}) 
\Omega_{\Dir} \,  \; 
\left| \frac{}{} \right. \; \xi_{j} \, \in 
\ms{N} , j= 1, \cdots ,n, \, \,  n  \in  \mbf{N}  \}  .
\end{equation}
In particular we simply call $\ms{F}_{\Dir}^{\fin}(\stateDirac )$ the finite
 particle subspace.
For $f \in \eltwo$ let us set 
\begin{align*}
&b^{\ast}_{1/2}(f) = B^{\ast}({}^{t} (f,0,0,0) ), \quad \quad  
b^{\ast}_{-1/2}(f) = B^{\ast}({}^{t}(0,f,0,0) ), \\
&d^{\ast}_{1/2}(f) = B^{\ast}({}^{t}(0,0,f,0) ), \quad \quad  
d^{\ast}_{-1/2}(f) = B^{\ast}({}^{t}(0,0,0,f) ) . 
\end{align*}
 Then it is seen that  
 \begin{align*}
&\{ b_{s} (f) , b_{\tau}^{\ast}(g) \} = 
\{ d_{s} (f) , d_{\tau}^{\ast}(g) \} =
\delta_{s, \tau} (f, g)_{\eltwo}   ,  \\
&\{ b_{s} (f) , b_{\tau} (g) \} = 
\{ d_{s} (f) , d_{\tau} (g) \} = \{ b_{s} (f) , d_{\tau} (g) \} = 
\{ b_{s} (f) , d_{\tau}^{\ast}(g) \} = 0  .
\end{align*}
   In this paper we denote the domain of operator $X$ by 
 $\ms{D} (X) $.
Let A be a self-adjoint operator on $\stateDirac $. 
The second quantization of $A$ is defined by 
\begin{equation}
\sqzf{A}  \;   
 =  \; \bigoplus_{n=0}^{\infty} \left( 
\sum_{j=1}^{n} ( I \tens \cdots I \tens 
\underbrace{ A}_{jth} \tens I   \cdots  \tens I )   \right) .
 \notag
\end{equation}
 Let $f \in \ms{D} (A)$.  Then it follows that 
\begin{align}
 &[ \sqzf{A}  \, ,  b_{s}(f)  ] \; = \;    - b_{s} (Af) , \qquad
 [ \sqzf{A}  \, ,  b_{s}^{\ast}(f)  ] \; = \;    b_{s}^{\ast} (Af) , 
  \label{dgammafb}  \\
 &[ \sqzf{A}  \, ,  d_{s}(f)  ] \; = \;  -   d_{s} (Af) , \qquad
 [ \sqzf{A}  \, ,  d_{s}^{\ast}(f)  ] \; = \;    d_{s}^{\ast} (Af) ,
 \label{dgammafd}
\end{align}
on the finite particle subspace. Now let us define the Dirac field.
  The energy of an electron with momentum $\mbf{p}$ 
 is given by
\begin{equation}
\qquad 
E_{M} (\mbf{p} )= \sqrt{ M^{2} + \mbf{p}^{2} } , \quad \quad M > 0 ,   \label{Diracmass}
\end{equation}
where the constant $M>0$ denotes the mass of an electron.
The free Hamiltonian of  the Dirac field is given  by
\begin{equation}
H_{\Dirac} \; =  \; \sqzf{E_{M}} . 
\end{equation}
Let  
\[
h_{D} (\mbf{p}) = \mbf{\alpha} \mbf{\cdot}
  \mbf{p}   + \beta M  , 
\qquad  \quad s(\mbf{p}) =  \mbf{s} \mbf{\cdot} \mbf{p} ,
\]
where  $\alpha^{j} $, $j=1,2,3$, and  $\beta$ are $4\times 4$ matrices satisfying the canonical anti-commutation relations 
\begin{equation}
 \{ \alpha^{j} , \alpha^{l} \} = 2 \delta_{j,l} , \qquad 
\{ \alpha_{j} , \beta \} = 0 ,   \qquad
\beta^{2}=I  , 
\end{equation} 
and $\mbf{s} = (s_{j})_{j=1}^3$ is the angular momentum of the spin. 
Let
  \begin{equation}
f_{s}^{l} (\mbf{p}) = 
\frac{\chi_{\Dirac}(\mbf{p} ) u_{s}^{l} (\mbf{p}) }{\sqrt{\piE (\mbf{p})}}     , \qquad  
g_{s}^{l} (\mbf{p})= 
\frac{\chi_{\Dirac} (\mbf{p}) \tilde{v}_{s}^{l} (\mbf{p})}
{\sqrt{\piE (\mbf{p} )}} , \notag 
\end{equation}
  where $ \chi_{\Dirac}$ is a cutoff function and, 
  $u_{s}= (u_{s}^{l})_{l=1}^{4} \; $ and $v_{s}= (v_{s}^{l})_{l=1}^{4} $, 
   $s= \pm 1/2 $,   denote the positive and negative energy part with
 spin $s $, respectively, satisfying 
 \begin{align*}
&h_{D} (\mbf{p}) u_{s} (\mbf{p}) = E_{M} (\mbf{p}) u_{s} (\mbf{p}), 
\qquad    s(\mbf{p} ) u_{s} (\mbf{p}) = s | \mbf{p} | u_{s} (\mbf{p}),   \\ 
&h_{D} (\mbf{p}) v_{s} (\mbf{p}) = 
-E_{M} (\mbf{p}) v_{s} (\mbf{p}), 
\qquad s(\mbf{p} ) v_{s} (\mbf{p}) = s | \mbf{p} | v_{s} (\mbf{p}) ,
\end{align*}
 and  we set $ \tilde{v}_{s}^{l} (\mbf{p})= v_{s}^{l} (-\mbf{p})$. 
   The field operator 
$\; \psi (\mbf{x}) \, = \, {}^{t}( \psi_{1}(\mbf{x}), \cdots , \psi_{4}(\mbf{x}) ) $
   is defined by 
\begin{equation}
\psi_{l}(\mbf{x}) = \sum_{s=\pm 1/2}( b_{s} (f_{s,\mbf{x}}^{l} ) +   d^{\ast}_{s} (g_{s, \mbf{x}}^{l} )) ,  \notag 
\end{equation}
where $\;   f_{s,\mbf{x}}^{l} (\mbf{p}) = f_{s}^{l} (\mbf{p}) 
e^{-i \mbf{p} \cdot \mbf{x}} \; $  and   
  $ \;  g_{s,\mbf{x}}^{l} (\mbf{p}) = g_{s}^{l} (\mbf{p}) 
e^{-i \mbf{p} \cdot \mbf{x}} \;$. 
We suppose the  assumption \textbf{(A.1)} below.  
\begin{quote}
\textbf{(A.1) (Ultraviolet cutoff for the Dirac field) }  \\  
$\quad $ $\chi_{\Dirac}$ satisfies that 
\[
\int_{\Rthree} \left| \frac{\chi_{\Dirac} (\mbf{p}) 
  u_{s}^{l} (\mbf{p})}{  \sqrt{E_{M}(\mbf{p}) }} \right|^{2} d \mbf{p}  \; <  \; \infty  , \qquad
\int_{\Rthree} \left| \frac{\chi_{\Dirac} (\mbf{p}) 
  \tilde{v}_{s}^{l} (\mbf{p})}{  \sqrt{E_{M}(\mbf{p}) }} \right|^{2} d \mbf{p}  \; <  \; \infty .
\]
\end{quote}
Since $b_{s} (f) $ and $d_{s} (f) $  are bounded 
 with 
\begin{equation} 
 \| b_{s} (f)  \|  = \| d_{s} (f)  \| = \| f \| ,  \label{boundbd}
 \end{equation}
 we  see that $\; \psi (\mbf{x}) \, = \, {}^{t}( \psi_{1}(\mbf{x}), \cdots , \psi_{4}(\mbf{x}) ) $ is bounded with
\begin{equation}
\| \psi_{l} ( \mbf{x} ) \| \leq M^{l}_{\, \D}, \quad \quad  \quad   l=1,\cdots , 4 ,
\label{boundpsil}
\end{equation}
$\quad $ \\ 
where $ \; M^{l}_{\, \D } = 
\sum\limits_{s=\pm 1/2} 
\left(  \frac{}{} \right. 
 \left\| \frac{\chi_{\Dirac} u_{s}^{ l}}{ \sqrt{(2 \pi)^{3} E_{M}}} 
\right\|
+  \left\| \frac{\chi_{\Dirac} \tilde{v}_{s}^{l }}{ \sqrt{(2 \pi)^{3} E_{M}}} \right\|
 \left. \frac{}{} \right) $.

$\quad$ \\

%-------------------------------------------------
%-------------------------------------------------

\subsection{Quantized Radiation Fields}
Next let us introduce the radiation field quantized in the Coulomb gauge. 
The Hilbert space for the quantized radiation field is given by
\begin{equation}
\ms{F}_{\rad} = \oplus_{n=0}^{\infty}
 (  \nstens L^{2} ( \Rthree ; \mbf{C}^{2}) ) ,  \notag 
\end{equation}
where $\nstens$ denotes the n-fold symmetric tenser
 product with $\tens_{s}^{0} \stateRad := \mbf{C}$.  
We denote the creation operator on $\ms{F}_{\rad} $ 
by $A^{\ast} ( \xi ) $, 
$\; \xi = (\xi_{1}, \xi_{2}) \in  L^{2} ( \Rthree ; \mbf{C}^{2}) ) 
$,  and the annihilation operator by $A ( \eta ) $,   
$\; \eta = (\eta_{1}, \eta_{2}) \in  
L^{2} ( \Rthree ; \mbf{C}^{2}) $.  Let $\Omegarad = \{ 1, 0, 0 \cdots \}  \; \in \; \ms{F}_{\rad} $ be
 the Fock vacuum.  
The finite particle subspace on $\ms{D} \subset  
\ms{F}_{\rad}$ is defiend  by 
 \[
\ms{F}_{\rad}^{\fin}(\ms{D} ) \; = 
 \; \textrm{L.h} \{  A^{\ast} (\xi_{1}) \cdots  A^{\ast} (\xi_{n})
\Omega_{\rad} \;| \; \xi_{j} \, \in 
\ms{D} , j= 1, \cdots ,n, \, \,  n  \in  \mbf{N} \}  .
\]
For simplicity we call
 $\ms{F}_{\rad}^{\fin}(  L^{2} (\Rthree ; \mbf{C}^{2}) )$ the finite particle subspace. 
The creation operator and the annihilation operator satisfy the canonical commutation relation on the finite particle subspace : 
\[
  [ A(\xi ) , A^{\ast} (\eta )] \;  = \;  ( \xi , \eta  )_{L^{2} (\Rthree ; \mbf{C}^{2})} \; , \quad \quad   [ A(\xi ) , A (\eta )  ] \; = \; [ A^{\ast}(\xi ) , A^{\ast} ( \eta ) ]  = 0 , 
\]
where $[X,Y] = XY -YX $.  
 For $f \in \eltwo $ let us set
\[
a^{\ast}_{1} (f) \; = \; A^{\ast}( (f,0) ) , \qquad 
a_{2}^{\ast} (f) \; = \; A^{\ast}((0,f)) .
\]
Then it follows that on the finite particle subspace 
\[
    [ \, a_{r}(f ), \, a^{\ast}_{r'} (g )  ] = \delta_{r,r'} (f, g  ), \qquad
    [ \, a_{r}(f ), \, a_{r'}(g )  ] = [ a_{r}^{\ast}(f ), \,  a_{r'}^{\ast}(g ) ] =0 .
\]
Let $S$ be a self-adjoint operator on $\stateRad $. The second quantization of $S$ is defined by 
\begin{equation}
\sqzb{S} \; 
 =  \; \bigoplus_{n=0}^{\infty} \left( 
\sum_{j=1}^{n} ( I \tens \cdots I \tens 
\underbrace{S}_{jth} \tens I   \cdots  \tens I )   \right) .
 \notag
\end{equation}
Let $f \; \in  \ms{D} (S^{-1/2} )$.
 It is seen that $ a_{r}(f)  $ and $ a_{r}^{\ast}(f) $ are relatively bounded with respect to $\sqzb{S}$ with
\begin{align}
& \| a_{r}(f) \Psi  \| \leq \| \frac{f}{\sqrt{\omega}} \| \, 
\|  \sqzb{S}^{1/2} \Psi \| , \label{boundar} \\
& \| a_{r}^{\ast}(f) \Psi  \| \leq \| \frac{f}{\sqrt{\omega}} \| \, 
\|   \sqzb{S}^{1/2} \Psi \| + \| f \| \| \Psi \| ,
\label{boundadr}
\end{align}
for $\Psi \in \ms{D} (\sqzb{S}^{1/2}) $. 
We also see that  
\begin{equation}
 [ \sqzb{S}  \, ,  a_{r}(f)  ] \; = \;  -  a_{r} (Sf) , \qquad
 [ \sqzb{S}  \, ,  a_{r}^{\ast}(f)  ] \; = \;    a_{r}^{\ast} (S f) , 
  \qquad   f \in \ms{D}(S) , 
 \label{dgammaa}
\end{equation}
on the finite particle subspace. \\ 
Let us define the quantized radiation field.
The one particle energy of  photon with momentum $\mbf{k}$ is given  by 
\begin{equation}
\omega (\mbf{k} ) \; =  \;  | \mbf{k} | .\label{photon_energy}
\end{equation}
Then the free Hamiltonian of the radiation field is given by 
\begin{equation}
 H_{\rad}  \;  = \; 
 \sqzb{\omega} .  
\end{equation}
Let 
\begin{equation}
h_{r}^{j}(\mbf{k}) = 
\frac{\chi_{\rad} (\mbf{k}) \varepsilon^{j}_{r}(\mbf{k}) }{\sqrt{2(2\pi )^{3} \omega (\mbf{k})}}   ,
\end{equation}
where $\chi_{\rad}$ is a ultraviolet cutoff function and 
 $\mbf{\varepsilon}_{r} (\mbf{k}) = 
( \varepsilon^{j}_{r} (\mbf{k}))_{j=1}^{3} $, $r=1,2$, denotes the polarization vectors satisfying  
 \begin{equation}
 \varepsilon_{r}(\mbf{k} ) \cdot \varepsilon_{r'}(\mbf{k} )=
 \delta_{r,r'} , \quad  
\mbf{k} \cdot  \varepsilon_{r}(\mbf{k} ) =0, \quad 
\text{a.e.} \;  \mbf{k} \in \Rthree.  \notag 
\end{equation}
It is seen in \cite{PhotonsAtoms} that any polarization vectors satisfy that 
\begin{equation}
\sum_{r=1,2}  \varepsilon_{r}^{j} ( \mbf{k} ) 
\varepsilon_{r}^{l} ( \mbf{k} ) \;   = \; 
  \delta_{j,l} \, - \frac{k^{j} k^l }{|\mbf{k}|^{2}} . \label{polarjl}
\end{equation}
 We introduce the following conditions. 
\begin{quote}
\textbf{(A.2) (Ultraviolet cutoff for the radiation field)     }  \\  
 $\quad$ $\chi_{\rad}$ satisfies that $ \; \overline{\chi_{\rad} (-\mbf{k}) )} \; = \; \chi_{\rad} (\mbf{k}) $ and
\[
\int_{\Rthree} \left| \frac{\chi_{\rad} (\mbf{k}) }{\sqrt{\omega (\mbf{k}) }} \right|^{2} d \mbf{k}  \; <  \; \infty 
,  \qquad  \qquad
\int_{\Rthree} \left| \frac{\chi_{\rad} (\mbf{k}) }{
\omega (\mbf{k}) } \right|^{2} d \mbf{k}  \; <  \; \infty .
\]
\end{quote}
The quantized radiation field $\; \mbf{A}(\mbf{x}) \, = \, 
 ( A^{j}(\mbf{x}) )_{j=1}^{3} \;  $ is defined by 
\begin{equation}
A^{j}(\mbf{x}) = \sum_{r=1,2}( a_{r}( h_{r,\mbf{x}}^{j}) +
 a^{\ast}_{r}( h_{r, \mbf{x}}^{j}))  , \qquad \mbf{x} \in \Rthree ,   
\end{equation} 
where $ h_{r,\mbf{x}}^{j}  (\mbf{k})  =
 h_{r}^{j}  (\mbf{k}) e^{-i \mbf{k} \cdot \mbf{x}} $. 
It is seen that   $ \textbf{A} (\mbf{x} ) $ is relatively bounded with respect to $ H_{\rad}^{1/2}  $   
 \begin{equation}
\| A^{j}(\mbf{x} ) \Psi \| \leq
\sum_{r=1,2} ( 2 M^{\, 2 ,j,  \, r}_{\R} \| H_{\rad}^{1/2} \Psi \| + 
M^{\, 1, j, \, r}_{\R} \| \Psi \|  ) , \label{boundAj}
\end{equation}
where 
$\;  M_{\R}^{\, l,j,r} \, = \,  \frac{1}{\sqrt{2 (2\pi )^{3}} } \left\|  \frac{ \chi_{\rad} \epsilon_{r}^{j} }{\sqrt{\omega}^{l}}    
  \right\|  $,
  $ \; k=1,2, \; r=1,2 $,  $  \; j=1,2,3$. \\

%-------------------------------------------------%------------------------------------------------

\subsection{Total Hamiltonian}
The total Hilbert space of quantum electrodynamics 
 is defined by
\begin{equation}
\ms{F}_{\QED} = \ms{F}_{\Dirac} \tens \ms{F}_{\rad} , \notag 
\end{equation}
 and the free Hamiltonian on  $ \ms{F}_{\QED}$ by 
 \begin{equation}
 H_{0} = H_{\Dirac} \tens I + I \tens H_{\rad} . \label{freeH}
 \end{equation}
 To define the interaction, we 
introduce an assumption on the 
 spatial cutoff functions $\chi_{\spa}$ : 
\begin{quote}
\textbf{(A.3)}  \textbf{(Spatial cutoffs)}  \\ 
 $\quad$ $\chi_{\spa}$ satisfies that 
\[  (i) \quad 
\int_{\Rthree} | \chi_{\spa} (\mbf{x}) | \dx < \infty  \qquad  
 and  \qquad (ii) \quad 
 \int_{\Rthree  \times \Rthree } \frac{| \chi_{\spa}
(\mbf{x} )  \; \chi_{\spa}(\mbf{y})| }{|\mbf{x} -\mbf{y}|}  
\dx \dy < \infty .
\]
\end{quote}

$\quad $ \\
If $ \chi \in L^{6/5} (\Rthree)$,  the  Hardy-Littlewood-Sobolev inequality  
 (e.g. \cite{LiLo}; 4.3 Theorem)  shows that $\chi$ 
 satisfies  the condition $(ii)$ in $\textbf{(A.3)}$.

$\quad $ \\ 
Now let us define the interaction. 
The electromagnetic current is denoted by 
\[
   J (\mbf{x}) \; \; = \; \; 
\left[
\begin{array}{c}
\rho (\mbf{x}) \\ \mbf{J} (\mbf{x})
\end{array}
\right] , 
\]
where $\; \rho(\mbf{x}) =\psi^{\ast} (\mbf{x}) \psi(\mbf{x})  $ and  $\; J^j (\mbf{x}) = \psi^{\ast} (\mbf{x})
 \alpha^j \psi (\mbf{x})  $,  $\; j=1, 2, 3$, with $ \alpha^j \in M_{4}(\mbf{C})$ satisfying $\{ \alpha^j  , \alpha^l \} = 2 \delta_{j,l}$.
Let  us define the functional on 
  $ \ms{D} (I \otimes H_{\rad}^{1/2} ) \, \times \, \ms{F}_{\QED} $ by 
 \[
\ell_{\I} ( \Psi , \Phi ) = \sum_{j=1}^{3} \int_{\Rthree} \chi_{\spa}(\mbf{x} ) 
( \current{\mbf{x}}{j} \otimes A^{j}(\mbf{x} ) \Psi, \, \Phi )_{\ms{F}_{\QED}} \dx  
\]
for $\Psi \in \ms{D} (I \otimes H_{\rad}^{1/2} ) $ and   
 $\Phi \in \FQED $. 
   By (\ref{boundAj}), (\ref{boundpsil}) and \textbf{(A.3)}, we  see that 
\begin{equation}
| \ell_{\I} (\Psi, \Phi ) | 
 \leq   \left(  L_{\I}  \| (I \otimes H_{\rad}^{1/2} ) \Psi \| + R_{\I} \| \Psi \| \right) \| \Phi \|  ,   \label{ell_one_bound}
\end{equation}
where
\begin{equation}
L_{\I} = 2 \| \chi_{\spa} \|_{L^{1}} \sum_{j, l, l',r } 
| \alpha_{l, \, l'}^{j} | \, 
\, M^{\, l}_{\R} \, M^{\, l'}_{\R} \, M^{\, 2 ,j,  \, r}_{\R} , 
\quad R_{\I} = \| \chi_{\I} \|_{L^{1}} \sum_{j, l, l',r } 
| \alpha_{l, \, l'}^{j} | \, 
\, M^{\, l}_{\R} M^{\, l'}_{\R} \, M^{\,1 ,j,  \, r}_{\R} . \quad 
\label{LR_one}
\end{equation}
  By the Riesz representation theorem, there exists a unique vector $ \, \Xi_{\Psi}  \in  \ms{F_{\QED}} \, $ 
such that 
\[
 \, \ell_{\I}( \Psi, \Phi )= ( \Xi_{\Psi} , \Phi ) \qquad
 \text{ for all } \quad \Phi \in \ms{F_{\QED}}  .
\]
 Let us define the linear operator $H'_{\I} \, : \, \ms{F}_{\QED} \; \to  \ms{F}_{\QED} $ by 
\begin{equation}
H'_{\I} : \Psi \longmapsto \Xi_{\Psi} . \label{def_interactI}
\end{equation}
It is seen from (\ref{ell_one_bound}) that
\begin{equation}
 \| H'_{\I}   \Psi \| \leq 
  L_{\I}  \| (I \otimes H_{\rad}^{1/2} ) \Psi \| + R_{\I} \| \Psi \| . \label{InteractboundI}
\end{equation}
We may express  $H'_{\I}$ formally by 
\[
H'_{\I} = \sum_{j=1}^{3} \int_{\Rthree} 
\,   
\chi (\mbf{x}) J^j (\mbf{x}) \tens
 A^{j}(\mbf{x} )   \,  \dx \, .
\]

$\quad $  \\ 
In a  similar way to   $H_{\I}'$, let us define  the functional  
 $\ell_{\II} : \ms{F_{\Dirac}} \times   \ms{F_{\Dirac}} \to \mbf{C}  $ by 
 \[
 \ell_{\II} (\Psi , \Phi ) =
 \int_{\Rthree \times \Rthree} 
\frac{\chi_{\spa}(\mbf{x} )  \chi_{\spa}(\mbf{y})}{|\mbf{x} - \mbf{y}|}
  ( \density{\mbf{x}} \density{\mbf{y}}
  \Psi, \Phi )_{\ms{F}_{\Dirac}} \dx \, \dy .
 \]
 By (\ref{boundpsil}) and  \textbf{(A.3)},  we see that
\begin{equation}
 | \ell_{\II} ( \Psi , \Phi ) | 
 \leq \left( M_{\II} \sum_{l, \nu , }
 ( M^{l}_{\, \D}  M^{\nu}_{\, \D} )^{2} \right) 
 \| \Psi \|    \|  \Phi \| ,    \label{ell_two_bound}
\end{equation} 
where  $ M_{\II} :=\int_{\Rthree  \times \Rthree } \frac{| \chi_{\spa}
(\mbf{x} )  \; \chi_{\spa}(\mbf{y})| }{|\mbf{x} -\mbf{y}|}  
\dx \dy $. 
 Then by using  the Riesz representation theorem again,
 it is seen that  there exists a unique vector $\Upsilon_{\Psi} \in \ms{F_{\Dirac }}  $ such that   for all 
$  \Phi \in \ms{F_{\Dirac }} $,
 \[
  \ell_{\II}(\Psi, \Phi ) = ( \Upsilon_{\Psi} , \Phi )
.\]
 Then we can define the linear operator $H_{\longti} : \ms{F}_{\Dirac} 
 \to \ms{F}_{\Dirac} $ by
\begin{equation}
 \Hlongti : \Psi \longmapsto \Upsilon_{\Psi}  .
  \label{def_V_Dirac}
\end{equation}
$\Hlongti$  is expressed by  formally
\[
 \Hlongti
=\int_{\Rthree \times \Rthree} 
\frac{\chi_{\spa}(\mbf{x} )  \chi_{\spa}(\mbf{y})}{|\mbf{x} - \mbf{y}|}
\density{\mbf{x}} \density{\mbf{y}}
    \dx \, \dy .
\]
Physically $\Hlongti$ derives from the longitudinal photons 
 \cite{BD}.  
By (\ref{ell_two_bound}), it is seen that $\Hlongti$ is bounded with 
\begin{equation}
\| \Hlongti \| \leq 
 M_{\II} \sum_{l, \nu }
 ( M^{\, l}_{\, \D}  M^{ \, \nu}_{\, \D} )^{2} .
 \label{bound_V_Dirac}
\end{equation}
Let 
\begin{equation}
H_{\II}' \; = \;   \Hlongti \tens I . \label{def_interactII}
\end{equation}
 $\quad $ \\ 
Then the total Hamiltonian is given by 
\begin{equation}
 H\;  = \;  H_{0} + 
e H'_{\I} + \frac{e^2}{8\pi}  H'_{\II} ,  \label{Total_H}
 \end{equation}
with the coupling constant $e \in \mbf{R}$. 
 On the  self-adjointness of $H$,  the following Lemma follows.

\begin{quote} 
\textbf{ Lemma} (\cite{Ta09}, Lemma 1.1 )
Assume that \textbf{(A.1)}-\textbf{(A.3)} hold. Then
 $H$ is self-adjoint on $\domain{\Hzero} $, and 
  essentially self-adjoint on any core of $H_{0}$ and bounded from below.
\end{quote}
 In particular $H$ is essentially self-adjoint on 
\begin{equation}
\ms{D}_{0} \; = \; \core  ,  \label{def_core}
\end{equation} 
where $\hat{\tens}$ denotes the algebraic tensor product.

$\quad$ \\
The scaled QED-Hamiltonian is defined  by  
\begin{equation}
H (\Lambda ) \;  = \; \HDirac \tens I \; 
 +   \; \Lambda^{2} I \tens \Hrad  \; + \;
e \Lambda H'_{\I}   \; + \;  \frac{e^2}{8 \pi}  H'_{\II}
 , \label{def_H_Lambda}
\end{equation}
where $\Lambda > 0 $ denotes the scaling parameter. 
 We are concerned with the asymptotic behavior of 
 $H (\Lambda ) $ as $\Lambda \to \infty $.  
The strategy is that  
we use the dressing transformation defined in
 (\ref{HzeroK}),  and take the scaling limit of the unitary transformed Hamiltonian.  
 The following theorem is the main result in this paper.
\begin{theorem}
 Assume \textbf{(A.1)}-\textbf{(A.3)}. Then for $z  \in 
  \mbf{C} \backslash \mbf{R}  $
\begin{equation}
s-\lim_{\Lambda \to \infty }
\left( H(\Lambda ) -z \frac{}{} \right)^{-1} \; \\  = 
\left( \HDirac + \frac{e^2}{8 \pi} \Hlongti  - \frac{e^2}{4}   \Veff \;   -z \right)^{-1}      \tens \POmegarad ,  \label{mainTheorem}
\end{equation}
where
\begin{equation}
\Veff \; =\; 
    \sum_{j,l}\int_{\Rthree \times \Rthree} 
  \;  \; \mbf{J}_{\chi} (\mbf{x}) \, \cdot \btriangle ( \mbf{x} -\mbf{y} ) 
 \mbf{J}_{\chi} (\mbf{y})   \; \; \dx \dy ,   \label{def_V_eff}
\end{equation}
where 
 $\; \btriangle (\mbf{z} ) =
 ( \lambda^{j,l} (\mbf{z}) + \lambda^{j,l} (\mbf{-z})   )_{j,i=1}^{3} $ is the $3 \times 3 $ matrix, and 
$\lambda^{j,l} (\mbf{z} )$ is a function 
  defined in (\ref{lambdajl}) . 
\end{theorem}

$\quad$ \\ 
By the general theorem (\cite{Su07-2}, Lemma 2.7) on resolvent convergence, we obtain
 the following corollary.
 \begin{corollary}
 Assume \textbf{(A.1)}-\textbf{(A.3)}. Then
\begin{equation}
s-\lim_{\Lambda \to \infty } e^{ -it H (\Lambda ) }
  (I \tens   \POmegarad ) \;   = 
  e^{ -it 
\left( \HDirac + \frac{e^2}{8\pi} \Hlongti - \frac{ e^2}{4} \Veff 
   \right) } 
  \tens \POmegarad  .
\end{equation}
\end{corollary}

%-------------------------------------------------%----------------------------------------------

%-------------------------------------------------%-------------------------------------------------

\section{Proof of Theorem 2.1}
To prove the Theorem 2.1, we apply the abstract scaling limit considered in \cite{Ar90}. \\ 
Let $\ms{X}$ and $\ms{Y}$ be Hilbert spaces. 
Let us set 
\[
\ms{Z} = \ms{X} \tens   \ms{Y} .
\]  
Let $A$ and $B$ be non-negative self-adjoint
 operators on $\ms{X}$ and $\ms{Y}$, respectively, 
and we assume ker $B \ne \{ 0 \} $. 
Let $P_{B} \;  : \ms{Y} \to \text{ ker } B $ the orthogonal
 projection. We consider a family of symmetric operators
 $\{ C(\Lambda ) \}_{\Lambda >0 } $ satisfying the 
 conditions :    
\begin{quote}
  (\textbf{C.1}) For all $\epsilon >0 $ there exists a constant 
 $\Lambda (\epsilon ) >0 $ such that for all 
$\Lambda > \Lambda (\epsilon )$, \\
 $\ms{D} (A \tens I) \cap 
 \ms{D} (I \tens B)  \subset \ms{D} (C(\Lambda ))$, 
 and there exists $ b(\epsilon ) \geq 0 $ such that
 \[
 \| C (\Lambda ) \Xi \| \leq 
  \epsilon \| ( A \tens I  + \Lambda I \tens B ) \Xi \| 
   + b (\epsilon ) \| \Xi \| .
 \]
\end{quote}
\begin{quote}
(\textbf{C.2}) There exists a symmetric operator $C$ on $\ms{Z}$ 
 such that   
 $ \ms{D} \tens \text{ker } B   \subset \ms{D} (C) $ 
 and for all $z \in \mbf{C} \backslash \mbf{R} )$, 
\[
s-\lim_{\Lambda \to \infty} 
C (\Lambda ) (A\tens I + \Lambda I \tens B -z) 
= C (A-z)^{-1} \tens P_{B}.
\]  
\end{quote}
\textbf{Theorem A } (\cite{Ar90}, Theorem 2.1) \\ 
Assume (\textbf{C.1}) and (\textbf{C.2}). Then (i)-(iii) follows.\\
(i) there exists $\Lambda_{0} \geq 0$ such that for all 
 $\Lambda > \Lambda_{0} $, 
 \[
 X( \Lambda ) = A \tens I + \Lambda I \tens B 
 + C (\Lambda )
 \]is self-adjoint on $\ms{D}(A \tens I ) \cap \ms{D} 
(  I \tens B ) $ and uniformly bounded from below for 
 $\Lambda$, furthermore  $X(\Lambda )$ is essentially self-adjoint on any core of $A \tens I + I \tens B $. \\ 
 (ii) Let $X= A \tens I + ( I \tens P_{B} )  C  ( I \tens P_{B} ) $.
 Then $X$ is self-adjoint on $\ms{D} (A \tens I )$ and  bounded from below, and essentially self-adjoint on any core of $A \tens I$. \\
(iii) Let $z \in \bigcap_{\Lambda \geq \Lambda_{0}} 
 \rho (X(\Lambda )) \cap \rho (X) $ where $\rho (\ms{O}) $ denotes the resolvent set of an operator $\ms{O}$.  Then,  it follows that 
\[
s-\lim_{\Lambda \to \infty} (X(\Lambda ) -z)^{-1}
 = (X-z)^{-1} (I \tens P_{B}) . 
\]

$\quad$ \\ 
Now let us consider $H_{\QED}$. Let
\begin{equation}
 \Pi^{j}(\mbf{x}) = i \sum_{r=1,2}
 \left(  -a_{r} (\frac{h_{r,\mbf{x}}^{j}}{\omega}) 
  + a_{r} (\frac{h_{r,\mbf{x}}^{j}}{\omega} ) \right) . \label{def_Pi}
 \end{equation}
 In a similar way to $H'_{\I}$, we can define the operator 
 \begin{equation}
 T= \sum_{j=1}^{3} \int_{\Rthree} \cutoff{x}{} 
 \left( \frac{}{} \current{x}{j} \tens \opePi{x}{j} \right) \dx . \label{def_J_Pi}
 \end{equation}
 By the canonical commutation relations of 
$a_{r}(f)$ and $a_{r'}^{\ast}(g)$, we have 
 \begin{equation}
 [ \Pi^{j}(\mbf{x}),  \Pi^{l}(\mbf{y})  ] = 0 , 
 \label{Pij_Pil}
 \end{equation}
 follows.  By (\ref{polarjl}), we also  see that 
 \begin{equation}
 [ A^{j} (\mbf{x}) , \Pi^{l} (\mbf{y}) ] = 
\lambda^{j,l} (\mbf{x} -  \mbf{y})  , \label{Aj_Pil}
 \end{equation}
where
\begin{equation}
\lambda^{j,l} (\mbf{z}) 
\; = \; 
\intRthree \frac{|\cutoff{k}{rad}|^{2}}{(2\pi )^{3} | \mbf{k} |^2}
\left ( \delta_{j,l } -\frac{k^{j} k^{l}}{|\mbf{k}|^{2}} \right) 
 e^{-i\mbf{k} \cdot \mbf{z} } \dk . \label{lambdajl}
\end{equation}
 By (\ref{boundar}) and (\ref{boundadr}), it is seen that
\begin{equation}
\| \Pi^{j}(\mbf{x} ) \Psi \| \leq
\sum_{r=1,2} (  M^{4 ,j,  \, r}_{\R} \| H_{\rad}^{1/2} \Psi \| + 
M^{3, j, \, r}_{\R} \| \Psi \|  ) .\label{boundPij}
\end{equation} 
 By (\ref{dgammaa}), it is seen that 
 \begin{equation}
 [ \Pi^{j} (\mbf{x}) , H_{\rad}  ] = -i \, A^{j} (\mbf{x})  , 
 \label{commutepirad}
\end{equation}
  on the finite particle subspace, and we have  
\begin{equation}  
[ T, I \otimes H_{\rad} ]
 = -i H'_{\I}   \label{commuteTHrad}
 \end{equation}
 on $\ms{D}_{0}$. Let us define the unitary transformation
  $U(t) $ by 
\begin{equation}
\qquad \qquad 
 U(t) \; = \; e^{itT} ,
 \qquad t \in \mbf{R}
. \label{def_Dress}
\end{equation}
 \begin{lemma}  \label{TAYLOR}
 Assume \textbf{(A.1)} - \textbf{(A.3)}. Then  there exists $\theta_{j} (t)  \in   [ -|t| , \; |t| ]$, 
$j=1, \cdots, 4$,  such that   on $\ms{D}_{0}$,
\begin{align}
& (i) \quad  U(t)^{-1} (H_{\Dirac}  \tens I  ) U (t) = 
H_{\Dirac}  \tens I + 
(-it) U(\theta_{1} (t))^{-1} 
 [ T , \, H_{\Dirac}  \tens I  ] U(\theta_{1} (t)) ,
 \label{theta1}   \\ 
& (ii) \quad U(t)^{-1} (I \tens H_{\rad} ) U (t) 
= I \tens H_{\rad} - t H'_{\I} + \frac{it^{2}}{2} 
U(\theta_{2} (t))^{-1} [  T, \;  H'_{\I} ] U(\theta_{2} (t)) , 
   \label{theta2} \\
& (iii) \quad U(t)^{-1} H'_{\I} U(t)
 = H'_{\I} + (-it) U(\theta_{3} (t))^{-1} 
 [ T , \, H'_{\I}  ] U(\theta_{3} (t))   ,  
   \label{theta3} \\
  &(iv) \quad U(t)^{-1} H'_{\II} U(t)
 = H'_{\II} + (-it) U(\theta_{4} (t))^{-1} 
 [ T , \, H'_{\II}  ] U(\theta_{4} (t))   .  \label{theta4}
\end{align}
 \end{lemma}
\textbf{(Proof)}
Let us  only prove (ii).   Other cases  can be  proven in a similar  manner to (ii). 
Let $\Psi \in \Dzero$ and $\Phi \in \ms{F}_{\QED}$.
  We set 
\[
  F_{\Phi, \Psi} (t)= ( \Phi, U(t)^{-1} \, (I \otimes H_{\rad} ) \, 
  U (t) \Psi   ). 
\]
  By the strong differentiability of $U(t) \Psi$ with respect to $t$, 
 Taylor's theorem shows that   
 there exists $\theta_{2} (t) \in [ -|t|, \, |t| ]$ such that 
\[
F_{\Phi, \Psi} (t) = F_{\Phi, \Psi} (0)
 + \frac{t}{1!} F'_{\Phi, \Psi} (0) + 
 \frac{t^{2}}{2!} F''_{\Phi, \Psi} (\theta_{2} (t)) ,  
\]  
where $F' = \frac{dF}{dt} $ and $F''= \frac{d^2 F}{d t^2 } $. 
 By (\ref{commuteTHrad}),  we obtain (\ref{theta2}), 
 since $\Phi \in  \ms{F}_{\QED}$ is arbitrary.
 $\blacksquare$ \\

$\quad $ \\
By Lemma \ref{TAYLOR} we obtain the following corollary.
\begin{corollary} \label{8/14.2}
  Assume \textbf{(A.1)} - \textbf{(A.3)}. Then it follows that
 \begin{equation}
 U\left( \frac{e}{\Lambda} \right)^{-1}
  H (\Lambda ) U\left( \frac{e}{\Lambda}\right)
  \; = \; \tilde{H}_{0} (\Lambda )
    \; + \; K  (\Lambda ) ,  \label{HzeroK}
\end{equation}
where 
\begin{equation}
 \tilde{H}_{0} (\Lambda ) \; = \; 
 \left(  \HDirac + \frac{e^2}{8\pi} \Hlongti \right)  \tens I \, + \, \Lambda^{2} \Hrad ,
  \label{HzeroLamb}
\end{equation}
and
\begin{align}
K(\Lambda )
 &= -i \frac{e}{\Lambda}  U\left(\theta_{1} (
\frac{e}{\Lambda}  )\right)^{-1} 
 [ T , \, H_{\Dirac}  \tens I  ] U\left( \theta_{1} 
(\frac{e}{\Lambda}) \right)  + 
 \frac{i e^{2}}{2}  U \left( 
 \theta_{2}  ( \frac{e}{\Lambda}  )\right)^{-1} 
 [ T , H'_{\I} ] U\left( 
\theta_{2}  (\frac{e}{\Lambda}) \right) \notag \\
 &  -i e^{2} U\left( \theta_{3} (
\frac{e}{\Lambda}  ) \right)^{-1} 
 [ T , H'_{\I} ] U\left( \theta_{3}  (\frac{e}{\Lambda})\right)
  -i \frac{e^3}{8\pi \Lambda}  \, U\left( \theta_{4} (
\frac{e}{\Lambda}  )\right)^{-1} 
 [ T , H'_{\II} ] U\left( \theta_{4}  
(\frac{e }{\Lambda}) \right) . \label{KLamb}
\end{align}
\end{corollary}

$\quad $ \\
By Corollary \ref{8/14.2}, it follows that  for $z \in \mbf{C} 
 \backslash \mbf{R}  $, 
 \begin{equation}
 ( H (\Lambda )  \; - z )^{-1} \; \; = \; \; 
 U\left( \frac{e}{\Lambda} \right)
  \left( \tilde{H}_{0} (\Lambda ) + K (\Lambda ) -z 
 \frac{}{} \right)^{-1} \,  U\left( \frac{e}{\Lambda}\right)^{-1} .
 \end{equation}
 In the following proposition, we will prove that  
 $\tilde{H}_{0} (\Lambda )$  and $\;  K (\Lambda )$ satisfy 
 the condition \textbf{(C.1)} and  \textbf{(C.2)} 
 with  applying  $ \tilde{H}_{0} (\Lambda )$ to 
$X_{0} (\Lambda )$ and $K(\Lambda )$ to 
 $C(\Lambda )$ .
 
\begin{proposition} \label{8/14.1}
  Assume \textbf{(A.1)} - \textbf{(A.3)}. \\
 \textbf{(1)} $\; $ For $\epsilon > 0$, 
there exists $\Lambda (\epsilon) \geq 0 $ such that 
 for all  $  \Lambda > \Lambda (\epsilon) $,
\begin{equation}
\qquad \| K(\Lambda )  \Psi \| \; 
\leq \;  \epsilon \| \tilde{H}_{0} (\Lambda ) \Psi \| \,  +
  \,  \nu (\epsilon ) \| \Psi \|  , \qquad \qquad \Psi \in \ms{D}_{0} , 
  \label{6/17.2}
\end{equation}  
 holds, where $ \nu (\epsilon )$ is a constant independent of 
 $ \Lambda \geq \Lambda (\epsilon) $. \\ 
\textbf{(2)} $\; $
For all $ z  \in  \mbf{C} \backslash  \mbf{R} $, it follows that 
\begin{equation}
 s-\lim_{\Lambda \to \infty }
K(\Lambda ) 
\left( \tilde{H}_{0} (\Lambda ) -z \right)^{-1}
 \;   =   \; K ( H_{\Dirac} + \frac{e^2}{8 \pi} \Hlongti  -z   )^{-1} \tens \POmegarad  ,
 \label{6/17.5}
 \end{equation}
 where
 \begin{equation}
K \; = \; - \frac{i e^{2}}{2} [ T, \, H'_{\I} ] . \label{defK}
\end{equation}
\end{proposition}

$\quad$ \\ 
To prove Proposition \ref{8/14.1}, let us prove 
 Lemma \ref{bound_H_Dirac_J} - Lemma \ref{KLambdaK}.
 
\begin{lemma} \label{bound_H_Dirac_J}
Assume (\textbf{A.1})-(\textbf{A.3}).   Then it follows that 
\begin{equation}
\qquad   
\left\| \frac{}{}  [ \HDirac    , \, \current{\mbf{x}}{j}  ] \right\| 
 \; \leq \;  c^{j},    \qquad \qquad 
 \qquad \qquad \label{H_Dirac_J}
\end{equation}
where $ \; c^{j} = \frac{2}{\sqrt{2 \pi}^{3}}  \sum\limits_{l,l'} 
 \sum\limits_{s}  | \alpha_{l,l'}^{j} |   
\left( \frac{}{} \|  \sqrt{E_{M}} \chi_{\D} u_{s}^{l} \| 
 +  \|  \sqrt{E_{M}} \chi_{\D} \tilde{v}_{s}^{l} \| \right) $, and 
\begin{equation}
 \qquad  
\left\| \frac{}{}  [  \density{x}  \density{y}    , \, \current{\mbf{x}}{j}  ] \right\| 
 \; \leq \;  d^{j} , \qquad \qquad 
 \qquad \qquad \label{rho_xy_J}   
\end{equation}
where $ \;  d^{j}  = 4 \sum\limits_{l,l' \nu , \nu '} 
|\alpha^{j}_{l,l'}| \left( M_{\D}^{\nu} M_{\D}^{\nu ' } \right)^{2} M_{\D}^{l} M_{\D}^{l'} $.
\end{lemma}
(\textbf{Proof}) \\ 
By using $ [X, YZ ] = [X, Y  ]Z  + Y[X,Z] $, we see that
\[
[\HDirac  , \current{\mbf{x}}{j} ] = \sum_{l,l'}  \alpha_{l,l'}^{j} 
 \left(  \frac{}{} [\HDirac , \psi^{\ast}_{l} (\mbf{x}) ]
 \psi_{l'} (\mbf{x})  \; + \; 
\psi^{\ast}_{l} (\mbf{x}) [\HDirac , \psi_{l'} (\mbf{x}) ] \right) .
\]
By the commutation relations (\ref{dgammafb}) and 
(\ref{dgammafd}), we have
\begin{align*}
 &[\HDirac , \psi^{\ast}_{l} (\mbf{x}) ]
\; = \sum_{s} \left(  b_{s}^{\ast} (E_{M} f_{s, \mbf{x}}^{l} ) 
 -  d_{s} (E_{M} g_{s, -\mbf{x}}^{l}  )  \right) , \\
& [\HDirac , \psi_{l'} (\mbf{x}) ]
\; = \sum_{s} \left( - b_{s} (E_{M} f_{s, \mbf{x}}^{l'}  )
 +  d_{s}^{\ast} (E_{M} g_{s, -\mbf{x}}^{l'}  ) \label{6/4.2} \right) .
\end{align*}
Then  by (\ref{boundbd}) and (\ref{boundpsil}), we obtain (\ref{H_Dirac_J}). 
We  also see that 
\[
[\density{x}  \density{y}, \current{\mbf{x}}{j} ] = \sum_{l,l'}  \alpha_{l,l'}^{j} 
 \left(  \frac{}{} [\density{x}  \density{y} , \psi^{\ast}_{l} (\mbf{x}) ]
 \psi_{l'} (\mbf{x})  \; + \; 
\psi^{\ast}_{l} (\mbf{x}) [\density{x}  \density{y} , \psi_{l'} (\mbf{x}) ] \right) . \label{6/4.4}
\]
Since 
\begin{equation}
\left\| [ \density{x}  \density{y} , \psi^{\sharp}_{l} (\mbf{x}) ] 
\right\| 
 \leq 2  \| \density{x} \| \,  \| \density{y} \| \, 
\| \psi^{\sharp}_{l} (\mbf{x}) \|   ,
\end{equation}
where $ \; \psi^{\sharp}_{l} (\mbf{x}) = \psi_{l} (\mbf{x}) 
 \; $ or $\;  \psi^{\ast}_{l} (\mbf{x}) $,
it follows from (\ref{boundpsil}), that   
\begin{equation}
\| [ \density{x}  \density{y} , \psi_{l}^{\, \sharp} (\mbf{x}) ] \| 
 \; \leq \; 2 \sum_{\nu , \nu ' }
  \left( M_{\D}^{\nu} M_{\D}^{\nu ' } \right)^{2} M_{\D}^{l} .
  \label{6/4.5}
\end{equation}
Then, by (\ref{boundpsil}), (\ref{6/4.4}) and (\ref{6/4.5}),  we have 
(\ref{rho_xy_J}). $\blacksquare$

\begin{lemma} \label{T_H_V_Dirac}
 There exist $a_{j}>0$, $b_{j}>0$,  $j=1,2$, independent of $s$, such that  for $\Psi \in \Dzero$
 \begin{align}
 &(i) \quad \| U(s)^{-1}[T, H_{\Dirac} \tens I ] U(s) \Psi \|  \leq a_{1} \| I \otimes H_{\rad}^{1/2} \Psi \| 
  + b_{1} \| \Psi \| ,    \label{bound_T_HDirac}  \\ 
 &(ii) \quad \| U(s)^{-1}[T, H_{\II}' ]  U(s) \Psi \|
 \leq a_{2} \| I \otimes H_{\rad}^{1/2} \Psi  \| 
  + b_{2} \| \Psi \|   \label{bound_T_VDirac}. 
 \end{align}
 \end{lemma}
(\textbf{Proof}) \\ 
(i)
Let $\Psi \in \Dzero$ and  $\Phi \in \ms{F}_{\QED}$.  
 We see that 
 \[
 (\Phi, U(s)^{-1}[T, H_{\Dirac} \tens I ] U(s) \Psi ) 
= \int_{\Rthree} \cutoff{x}{}
  \left( \frac{}{} (   [  H_{\Dirac}  ,  \current{x}{j} ] \tens I ) U(s) \Phi ,    \;  ( I \tens  \opePi{x}{j} )  U(s) \Psi  \right) \dx . 
\]
 Note that $ [  H_{\Dirac}  ,  \current{x}{j} ] $
  is a  bounded operator by (\ref{H_Dirac_J}). 
Then, by the Schwarz inequality, we have  
 \begin{multline}
 | (\Phi, U(s)^{-1}  [T, H_{\Dirac}   \tens I  ]  U(s) \Psi ) | 
\\ 
\leq \left( \int_{\Rthree}  | \cutoff{x}{} |  \, 
 \| \left( \frac{}{} [ H_{\Dirac}  , 
 \current{x}{j} ] \tens I \right) U(s) \Phi \|^{2}  \dx 
  \right)^{1/2}  
  \times     \left( \int_{\Rthree} | \cutoff{x}{} | 
 \| ( I \otimes \opePi{x}{j} ) U(s) \Psi \|^{2} \dx  \right)^{1/2} . 
\label{6/4.6}     
\end{multline}
By (\ref{H_Dirac_J}),  we have 
\begin{equation}
 \left\| \left( [ \HDirac , \current{x}{j} ] \tens I \frac{}{} \right)U(s) \Phi  \right\| 
 \; \leq \;  c^{j} \| \Psi \| . \label{6/4.7}
\end{equation}
 Since $I \tens  \opePi{x}{j}$ and  $U(s) $ commute on 
 $ \ms{D}_{0}$
 for each $\mbf{x} \in \Rthree$, 
 we have by (\ref{boundPij}) that,
 \begin{equation}
\|  I \tens  \opePi{x}{j}  U(s) \Psi \| 
=  \|  I \tens  \opePi{x}{j}  \Psi \| 
 \leq  \sum_{r=1,2} \left(  M^{4 ,j,  \, r}_{\R} \| 
 I \tens H_{\rad}^{1/2} \Psi \| + 
M^{3, j, \, r}_{\R} \| \Psi \|  \frac{}{} \right) . \label{6/4.8}
 \end{equation}
 Then by  (\ref{6/4.7}), (\ref{6/4.8}) and  (\ref{6/4.6}), we see that there exist $a_{1} \geq 0 $ and 
$b_{1} \geq 0 $ such that 
\[
\left| \frac{}{} (\Phi, U(s)^{-1}[T, H_{\Dirac} \tens I ] U(s) \Psi ) 
 \right|   \; \leq \; 
  \left( a_{1} \| I \otimes H_{\rad}^{1/2} \Psi \| 
  + b_{1} \| \Psi \|  \right)  \| \Phi \| 
\]
for $\Phi \in \ms{F}_{\QED}$.  Hence we have 
\[
 \| U(s)^{-1}[T, H_{\Dirac} \tens I ] U(s) \Psi \| 
 \; \leq \; 
   a_{1} \| I \otimes H_{\rad}^{1/2} \Psi \| 
  + b_{1} \| \Psi \| .
\]
(ii)
 Let $\Psi \in \Dzero$ and  $\Phi \in \ms{F}_{\QED}$.
 Then we have
\begin{align*}
  &(\Phi, U(s)^{-1}[T, \Hlongti \tens I  ] U(s) \Psi ) \\  
 &=  \int_{\Rthree} \cutoff{x}{} 
 \frac{\cutoff{y}{} \cutoff{z}{} }{| \mbf{y}-\mbf{z} |} 
  ( U(s) \Phi, \;
 \left(  [  \current{x}{j} , \rho (\mbf{y})  \rho (\mbf{z} ) ]
  \tens \opePi{x}{j} \right)  U(s) \Psi  )  \dx  \dy \dz \\  
 & = \int_{\Rthree} \cutoff{x}{} 
 \frac{\cutoff{y}{} \cutoff{z}{} }{| \mbf{y}-\mbf{z} |} 
  ( ( [ \rho (\mbf{z} ) \rho (\mbf{y}) , \current{x}{j} ] \tens I ) 
U(s) \Phi, \; 
  ( I  \tens \opePi{x}{j} ) U (s) \Psi  )  \dx  \dy \dz .
\end{align*}
By the Schwarz inequality,  we see  that  
\begin{align}
 | (\Phi, U(s)^{-1} [T, H'_{\II} ] U(s) \Psi ) | 
 &\leq  \left(
 \int_{\Rthree} | \cutoff{x}{} |
 \frac{ | \cutoff{y}{} \cutoff{z}{} |}{| \mbf{y}-\mbf{z} |} 
 \| ( [  \rho (\mbf{z} ) \rho (\mbf{y}) , \current{x}{j}   ]  \tens I) 
  U(s) \Phi \|^{2} \dx  \dy \dz \right)^{1/2} \notag \\ 
  & \quad \times  \left(
 \int_{\Rthree} | \cutoff{x}{} |  
 \frac{ | \cutoff{y}{} \cutoff{z}{} | }{| \mbf{y}-\mbf{z} |} 
 \| (I \tens \opePi{x}{j} ) U(s) \Psi \|^{2} \dx  \dy \dz
 \right)^{1/2} . \label{6/4.9}
\end{align}
By  (\ref{rho_xy_J}), (\ref{6/4.8}) and (\ref{6/4.9}), there exists $a_{2} \geq 0 $ and 
 $b_{2} > 0 $ such that 
\[
| (\Phi, U(s)^{-1} [T, H'_{\II} ] U(s) \Psi ) | 
 \leq \left( a_{2} \| I \otimes H_{\rad}^{1/2} \Psi  \| 
  + b_{2} \| \Psi \|    \right) \|  \Phi \|  ,
\]
for $\Phi \in \ms{F}_{\QED}$.
Then we have
\[
 \| U(s)^{-1} [T, H'_{\II} ] U(s) \Psi  \| 
 \leq  a_{2} \| I \otimes H_{\rad}^{1/2} \Psi  \| 
  + b_{2} \| \Psi \| .
\]
Thus the proof is completed. 
$\blacksquare$

$\quad $ \\ 
It is seen  that 
$| \lambda^{j,l} (\mbf{z} ) | $
 is uniformly bounded with respect to $\mbf{z}$, namely 
\begin{equation}
| \lambda^{j,l} (\mbf{z})  | \;   \leq  \; 
\gamma^{j,l} \; :=  
\intRthree \frac{|\cutoff{k}{rad}|^{2}}{(2\pi )^{3} | \mbf{k} |^2}
\left| \left( \delta_{j,l } -\frac{k^{j} k^{l}}{|\mbf{k}|^{2}}
 \right) \right| 
  \dk  \;     .
 \label{upperlambdajl}
\end{equation}

\begin{lemma} \label{BOUND_T_HI}
Assume (\textbf{A.1}) - (\textbf{A.3}). Then  
there exist constants $a_{3}>0$ and $b_{3} >0$, independent of 
sufficiently small $s$, such that for $\Psi \in \ms{D}_{0}$,
\begin{equation}
\| U(s)^{-1}[T, H'_{\I} ] U(s) \Psi \|
 \leq a_{3} \| I \otimes H_{\rad}  \Psi \| 
  +  b_{3}  \,  \| \Psi \| .   \label{bound_T_HI}
\end{equation} 
\end{lemma}
\textbf{(Proof)}
Let $\Psi \in \Dzero$ and $\Phi \in \ms{F}$.  
By the equality $[A\tens B , C \tens D ]
= [A,B ] \tens CD  + CA\tens [B,D ] $ and 
 the commutation relations (\ref{Aj_Pil}), we have
\begin{align}
&(\Phi, U(s)^{-1} [T, H'_{\I}]  U(s) \Psi )   \notag \\ 
&= \sum_{j,l}\int_{\Rthree \times \Rthree} 
 \cutoff{x}{} \cutoff{y}{} 
 \left(  U(s) \Phi , [ \current{x}{j}  \otimes \opePi{x}{j} 
  ,  \current{y}{l}  \otimes \opeA{y}{l} ] U(s) \Psi \right)
   \dx \dy  \notag \\
& = \sum_{j,l} \left( \frac{}{} X_{j,l}(\Phi, \Psi ) + Y_{j,l}(\Phi, \Psi ) \right), 
\label{Xjl_Yjl}
 \end{align}
 where
 \begin{align}
 & X_{j,l} (\Phi, \Psi ) 
 =   \int_{\Rthree \times \Rthree} 
 \cutoff{x}{} \cutoff{y}{} 
 \left(  ( [ \current{y}{l} , \current{x}{j}   ] \tens I )  U(s) \Phi ,
 \, (I \tens \opePi{x}{j}  \opeA{y}{l} ) U(s) \Psi \right)
   \dx \dy \label{def_Xjl} , \\
&  Y_{j,l}(\Phi, \Psi )
= \int_{\Rthree \times \Rthree} 
 \cutoff{x}{} \cutoff{y}{} \lambda^{j,l} (\mbf{x} - \mbf{y})
 \left(  U(s) \Phi ,   ( \current{y}{l}  \current{x}{j}  
\otimes I ) U(s) \Psi \right) 
   \dx \dy  .  \label{def_Yjl}
 \end{align}
By the Schwarz inequality, 
 \begin{align}
| X_{j,l} (\Phi, \Psi )| 
&\leq
\left\{ \int_{\Rthree \times \Rthree} 
 | \cutoff{x}{} \cutoff{y}{} |
   \|  ( [ \current{y}{l} , \current{x}{j}    ] \otimes I )
   U(s) \Phi \|^{2} \dx \dy \right\}^{1/2} \notag \\
&\qquad \times
\left\{ \int_{\Rthree \times \Rthree} 
 | \cutoff{x}{} \cutoff{y}{} | \|  ( I \otimes \opePi{x}{j}  \opeA{y}{l} ) U(s) \Psi \|^{2} 
 \dx \dy \right\}^{1/2} . \label{bound_Xjl}
 \end{align}
We see that  $\|   \|  [ \current{y}{l} , \current{x}{j}    ] \| 
 \leq 2 \| \current{y}{l} \|  \| \current{x}{j} \| $, 
  and hence, we have  from (\ref{boundpsil}) that 
 \begin{equation}
 \|  [ \current{y}{l} , \current{x}{j}    ] \| 
   \; \leq \; 
   2 \sum_{\mu , \mu ' ,  \nu ,  \nu ' }
   |\alpha_{\mu ,\mu' }^{l} |  |\alpha_{\nu ,\nu' }^{j} |
   M_{\D}^{\mu}  M_{\D}^{\mu ' } 
   M_{\D}^{\nu}  M_{\D}^{\nu ' }  . 
   \label{6/11.1}
 \end{equation}
In a similar way to Lemma \ref{TAYLOR}, it is seen that 
there exists $\tau (s) \in [ -|s| , |s|] $ such that  for 
$ \Psi \in \ms{D}_{0} $
\begin{align}
& U(s)^{-1} \left( \frac{}{} I \otimes \opePi{x}{j}  \opeA{y}{l}
 \right)  U(s)  \Psi
 \notag  \\
&= \quad 
\left( I \otimes \opePi{x}{j}  \opeA{y}{l} \right) \Psi 
+ \frac{-is}{1!} U(\tau (s))^{-1} 
[T , I \otimes \opePi{x}{j}  \opeA{y}{l} ] U(\tau (s) ) \Psi .
 \label{U_Pi_A}
\end{align}
 Let $ \Xi \in \ms{F}_{\QED}$. Then by the commutativity 
  of $I \tens \opePi{x}{j}$  and $T $, 
we have 
\begin{align}
&( \Xi ,   [T , I \otimes \opePi{x}{j}  \opeA{y}{l} ] U(\tau (s) ) \Psi ) \notag \\
&= (  \Xi , I \otimes \opePi{x}{j}  [T , I \tens  \opeA{y}{l} ] U(\tau (s) ) \Psi ) \notag  \\ 
&= \int_{\Rthree} \cutoff{z}{} \lambda^{k,l} (\mbf{z} - \mbf{y}) 
 ( \Xi , \, 
[ \current{z}{k} \tens \opePi{x}{j} , I \tens \opeA{y}{l} ]
U ( \tau (s) ) \Psi ) \dz 
 \notag \\
&=   \int_{\Rthree} \cutoff{z}{} \lambda^{k,l} (\mbf{z} - \mbf{y}) 
 ( ( \current{z}{k} \tens I )  \Xi , \,   U ( \tau (s)  ) 
( I \tens \opePi{x}{j} )  \Psi ) \dz .
\end{align}
Then we obtain  by (\ref{boundPij}) that 
\begin{align}
&\left| \frac{}{} ( ( \current{z}{k} \tens I ) \Xi ,   [T , I \otimes \opePi{x}{j}  \opeA{y}{l} ] 
U(\tau (s) ) \Psi ) \right| \notag   \\
& \leq \int_{\Rthree} |  \cutoff{z}{}| \, | \lambda^{k,l} 
(\mbf{z} - \mbf{y}) |
 \| ( \current{z}{k} \tens I ) \Xi \|  \, \| 
 ( I \tens \opePi{x}{j} )  \Psi  \|  \dz  \notag \\
 & \leq 
 \| \chi_{}  \|_{L^{1}} 
 \gamma^{k,l} \sum_{\nu , \nu '} | \alpha^{k}_{\nu , \nu '} |
 M_{\D}^{\nu} M_{\D}^{\nu ' }   
\left(  \frac{}{} 
  \sum_{r=1,2} (  M^{4 ,j,  \, r}_{\R} \|
 ( I \tens  H_{\rad}^{1/2} ) \Psi \| + 
M^{3, j, \, r}_{\R} \| \Psi \| )  \frac{}{}  
 \right) \| \Xi \| ,
 \end{align}
 where $\gamma^{k,l}$ is defined in (\ref{upperlambdajl}).
Hence we obtain that
\begin{equation}
\|  U(\tau (s))^{-1} [T , I \otimes \opePi{x}{j}  \opeA{y}{l} ] 
U(\tau (s) ) \Psi   \|  \; 
  \leq  \; 
  q_{j,l}  \| (  I \tens  H_{\rad}^{1/2}  ) \Psi \| + 
  \tilde{q}_{j,l} \| \Psi \|  ,  \label{bound_U_Pi_Al_U}
\end{equation}
where
\begin{align*}
&q_{j,l}= \| \chi_{\spa}  \|_{L^{1}} 
 \sum_{k, r, \nu , \nu '} \gamma^{k,l} | \alpha^{k}_{ \nu , \nu '} |
 M_{\D}^{\nu} M_{\D}^{\nu ' }   
       M^{4 ,j,  \, r}_{\R}  ,  \\
&\tilde{q}_{j,l} = \| \chi_{\spa}  \|_{L^{1}} 
 \sum_{k, r, \nu , \nu '}  \gamma^{k,l} | \alpha^{k}_{\nu , \nu '} |
 M_{\D}^{\nu} M_{\D}^{\nu ' }  
        M^{3 ,j,  \, r}_{\R} .
\end{align*}
It is seen that   there exist $c_{j,l} \geq 0 $
 and  $d_{j,l} \geq 0 $ such that 
\begin{equation}
 \| I \tens \opePi{x}{j} \opeA{y}{l} \Psi \| 
 \leq c_{j,l}  \| I \tens \Hrad  \Psi \| \, + d_{j,l} \| \Psi \| .
 \label{boundPiAl}
 \end{equation}
 By  (\ref{bound_U_Pi_Al_U}), 
 (\ref{boundPiAl}) and   (\ref{U_Pi_A}),   we have
 \begin{equation}
 \| U(s)^{-1} ( I \tens \opePi{x}{j} \opeA{y}{l} ) U(s) \Psi \| 
 \leq (c_{j,l} + s q ) \| I \tens \Hrad \Psi \| 
  +  (d_{j,l}  + s \tilde{q} ) \| \Psi \|  .
  \label{6/11.2}
 \end{equation}
 Hence by applying (\ref{6/11.1}) and   (\ref{6/11.2}), 
 to  (\ref{bound_Xjl}),  we see that 
there exist constant $a \geq 0$ and  $b \geq 0$ independent  of   sufficiently small $s$, such that  
\begin{equation}
| X_{l,j} (\Phi, \Psi )  | \leq \left( \frac{}{}  a  \| I \tens  \Hrad  \Psi \| \,   +  b \| \Psi \|  \right) \| \Phi \| .
 \label{6/11.3}
\end{equation} 
Furthermore we can see by (\ref{boundpsil}) under  (\textbf{A.3}) that 
\begin{align}
| Y_{l,j} (\Phi, \Psi )  | & \leq 
 \int_{\Rthree \times \Rthree}
    | \cutoff{x}{} \cutoff{y}{} 
    \lambda^{j,l} (\mbf{x} - \mbf{y} ) | 
 \, | ( U(s) \Phi , \, (\current{y}{l}  \current{x}{j} \tens I ) 
  U (s ) \Psi  ) | \, \dx \dy \notag   \\ 
 & \leq \| \chi_{} \|^{2}_{L^{1}} 
 \gamma^{j,l}
  ( \alpha_{\nu \nu '}^{j}   \alpha_{\mu \mu '}^{j} 
  \sum_{\nu, \nu ' , \mu , \mu '}M_{\D}^{\nu} M_{\D}^{\nu ' } 
 M_{\D}^{\mu} M_{\D}^{\mu ' }  ) \| \Phi \| \, \| \Psi \| . 
  \label{6/11.4}
\end{align}
By  (\ref{6/11.3}), (\ref{6/11.4}) and  (\ref{Xjl_Yjl}),
 it can be  seen that  there exist $a_{3} \geq 0 $ and $b_{3} \geq 0 $ 
 independent of sufficiently small $s$ 
 such that 
\[
  \left| \frac{}{}  (   \Phi, U(s)^{-1}  [  T, H_{\I}' ] U(s) \Psi  )
   \right|  \leq 
   \left( a_{3} \| I \tens \Hrad \Psi \| + b_{3} \| \Psi \| \right) 
   \| \Phi \| ,
\]
and hence we obtain that 
\[
  \left| \frac{}{}  (   \Phi, U(s)^{-1}  [  T, H_{\I}' ] U(s) \Psi  )
   \right|  \leq 
   a_{3} \| I \tens \Hrad \Psi \| + b_{3} \| \Psi \|  .
\]
Thus the proof is completed. 
 $\blacksquare$

%-------------------------------------------------%-------------------------------------------------

\begin{lemma}
 \label{KLambdaK}
Assume (\textbf{A.1})-(\textbf{A.3}). Then 
\begin{equation}
 s-\lim_{\Lambda \to \infty } K (\Lambda ) \Psi 
 \; =  \;  K \Psi , \qquad \qquad \qquad \Psi \in  \ms{D}_{0}.  
\end{equation}
where $K $ is an operator defined in (\ref{defK}) .
\end{lemma}
(\textbf{Proof}) \\ 
By Lemma \ref{T_H_V_Dirac}, it is sufficient to prove that 
\begin{equation}
 \lim_{t \to 0} U(t)^{-1}[ T, H_{\I}' ] U (t) \Psi \;
  = [ T, H_{\I}' ]  \Psi ,
  \label{6/17.1}
\end{equation}
 for $\Psi \in \ms{D}_{0}$. 
In a similar way to  Lemma \ref{TAYLOR}, 
there exists $\theta (t) \in [ -t, t]$ such that 
\[
U(t)^{-1}[ T, H_{\I}' ] U (t) \Psi \; = 
 \; [ T, H_{\I}' ] \Psi - it U( \theta ( t))^{-1}
[ T, H_{\I}' ] U (\theta (t)) \Psi .
\]
Then by Lemma \ref{BOUND_T_HI},  we obtain that
\[
\| U(t)^{-1}[ T, H_{\I}' ] U (t) \Psi   -  \; [ T, H_{\I}' ] \Psi \|
\; \leq \; t \;  \left(  \frac{}{}
 a_{3} \| I \tens H_{\rad} \Psi \| + b_{3} \| \Psi \| \right) ,
\]
 and hence 
 (\ref{6/17.1}) follows. $\blacksquare$

$\quad$  \\
$\quad $ \\
\textbf{{\large (Proof of Proposition \ref{8/14.1})}} \\
\textbf{(1)} By Lemma \ref{T_H_V_Dirac} and 
 Lemma \ref{BOUND_T_HI}, (\ref{6/17.2}) follows. \\
\textbf{(2)}  
It is seen that
\begin{multline}
K(\Lambda ) 
\left( \tilde{H}_{0} (\Lambda ) - \frac{}{}z  \right)^{-1} \;
 \;  = \; K(\Lambda ) 
( H_{\Dirac}  + \frac{e^2}{8 \pi} \Hlongti - \frac{}{} z  )^{-1}
 \tens \POmegarad     \\
+ K(\Lambda ) 
\left( \tilde{H}_{0} (\Lambda ) - \frac{}{}z  \right)^{-1}  
 \left(  I \tens \POmegarad^{\bot} \frac{}{} \right) , \qquad
\end{multline} 
where $ \POmegarad^{\bot} = 1 - \POmegarad $. 
By Lemma \ref{KLambdaK},  we have 
\begin{equation}
s-\lim_{\Lambda \to \infty }
 K(\Lambda )  \left( \frac{}{}
( H_{\Dirac} + \frac{e^2}{8 \pi} \Hlongti   - \frac{}{} z  )^{-1}
 \tens \POmegarad  \right) \Psi  \; = K 
 \left( \frac{}{}
( H_{\Dirac} + \frac{e^2}{8\pi}  \Hlongti - \frac{}{} z  )^{-1}
 \tens \POmegarad \right) \Psi .   \label{6/17.3}
\end{equation}
By  (\ref{6/17.2}), we see that 
 for $\epsilon >0 $ there exists $ \Lambda (\epsilon ) \geq 0 $ such that for all $ \Lambda > \Lambda (\epsilon )$  
\begin{equation}
\|  K(\Lambda ) 
\left( \tilde{H}_{0} (\Lambda ) - \frac{}{}z  \right)^{-1}  \Xi \| 
\leq  \epsilon \| \Xi \| + 
( \epsilon |z|  + \nu (\epsilon  ) ) \|  \left( \tilde{H}_{0} (\Lambda ) - \frac{}{}z  \right)^{-1}  \Xi \| ,  \notag
\end{equation}
follows for $\Xi \in \ms{F}_{\QED}$. Furthermore we see that 
$
 \lim_{\Lambda \to \infty} \; \left\|  \left( \tilde{H}_{0} (\Lambda ) - \frac{}{}z  \right)^{-1}    \left(  I \tens  \POmegarad^{\bot}  \frac{}{} \right)  \Psi  \right\| \; = \; 0 $, 
and hence we obtain that 
\begin{equation}
\lim_{\Lambda \to \infty } \left\| 
 K(\Lambda ) 
\left( \tilde{H}_{0} (\Lambda ) - \frac{}{}z  \right)^{-1}  
 \left(  I \tens  \POmegarad^{\bot}  \frac{}{} \right)  \Psi  \right\|
  \; = \; 0  . \label{6/17.4}
\end{equation}
By (\ref{6/17.3}) and (\ref{6/17.4}), 
 we obtain (\ref{6/17.5}).  
$\blacksquare $

$\quad$ \\ 
\textbf{{\large (Proof of Theorem 2.1)}} \\
We see that for $z \in \mbf{C} \backslash \mbf{R}$,
\[
  \left( \frac{}{} H (\Lambda ) -z \right)^{-1}
  \; = \; U \left(  \frac{e}{\Lambda} \right)  \left( \frac{}{}  \tilde{H}_{0}(\Lambda ) 
+ K (\Lambda )  - z \right)^{-1}  U \left(  \frac{e}{\Lambda} \right)^{-1}.
\]
It is seen from (\ref{6/17.2}) and  (\ref{6/17.5}) that
 $ \tilde{H}_{0} (\Lambda )$ and $K(\Lambda )$ satisfy 
  \textbf{(C.1)} and \textbf{(C.2)}  with  applying  $ \tilde{H}_{0} (\Lambda )$ to 
$X_{0} (\Lambda )$ and $K(\Lambda )$ to 
 $C(\Lambda )$. 
Then by Proposition \ref{8/14.1}, we obtain that 
\[
s-\lim_{\Lambda \to \infty }
 \left(  \frac{}{} H(\Lambda ) -z  \right)^{-1}  \; 
  =  \;  \left( \frac{}{} ( H_{\Dirac} + \frac{e^2 }{8 \pi} H_{\II}' ) \tens I 
 +K_{\rad} -z \right)^{-1}  \; (I \tens \POmegarad ) ,
\]
where 
\[
K_{\rad } \;   = \;   \frac{-i e^2}{2} 
(I \tens \POmegarad) [T, H'_{\I}] (I \tens \POmegarad ) .
\]
 Let us compute $K_{\rad}$.
 It is seen that
 \begin{equation}
 ( \Omegarad , \opePi{x}{j} \opeA{y}{l} \Omegarad )
  =  \frac{-i}{2} \lambda^{j,l} (\mbf{x} - \mbf{y} ) .
\end{equation}
 By (\ref{Aj_Pil}),  we see that for  $\Psi = \Psi_{\Dir} \tens \Omega_{\rad} \; $ and  
 $\; \Psi_{\Dir} \in \ms{F}_{\Dirac}$, 
 \begin{align}
&(\Psi, [T, H'_{\I}]   \Psi )   \notag \\ 
&=\sum_{j,l}\int_{\Rthree \times \Rthree} 
 \cutoff{x}{} \cutoff{y}{} 
 \left(   \Psi , [ \current{x}{j}  ,  \current{y}{l}  ]
\otimes \opePi{x}{j}  \opeA{y}{l}   \Psi 
\right)_{\ms{F}_{\QED}}
   \dx \dy  \notag \\
  &\qquad + \sum_{j,l}\int_{\Rthree \times \Rthree} 
 \cutoff{x}{} \cutoff{y}{} 
 \left(  \Psi ,   \current{y}{l}  \current{x}{j}  
\otimes [ \opePi{x}{j} , \opeA{y}{l} ]  \Psi \right)_{\ms{F}_{\QED}}
   \dx \dy  \notag \\ 
&=-\frac{i}{2} \sum_{j,l}\int_{\Rthree \times \Rthree} 
 \cutoff{x}{} \cutoff{y}{} 
 \lambda^{j,l} (\mbf{x} - \mbf{y} )
 \left(   \Psi_{\Dir} , [ \current{x}{j}  ,  \current{y}{l}  ] 
  \Psi_{\Dir} 
\right)_{\ms{F}_{\Dirac}}
   \dx \dy  \notag \\
  &\qquad -i \sum_{j,l}\int_{\Rthree \times \Rthree} 
 \cutoff{x}{} \cutoff{y}{} \lambda^{j,l} (\mbf{x} - \mbf{y} )
 \left(  \Psi_{\Dir} ,   \current{y}{l}  \current{x}{j}  
\Psi_{\Dir} \right)_{\ms{F}_{\Dirac}}
   \dx \dy . 
    \end{align}
Thus we have
\[
(I \tens \POmegarad) [T, H'_{\I}] (I \tens \POmegarad )  \; = \; 
 -\frac{i}{2} \sum_{j,l}\int_{\Rthree \times \Rthree} 
 \cutoff{x}{} \cutoff{y}{} 
 \lambda^{j,l} (\mbf{x} - \mbf{y} )
 \left(   \current{x}{j}   \current{y}{l} 
   + \current{y}{l} \current{x}{j}  
\right)
   \dx \dy . 
\] 
Then  the theorem follows.  $\blacksquare$

$\quad$ \\
{\Large Acknowledgments} \\
It is pleasure to thank Professor  F. Hiroshima for his advice and comments.

\end{document}